\documentclass[oneside,english,reqno]{amsart} 
\usepackage[T1]{fontenc}
\usepackage[latin9]{inputenc}
\usepackage{amsmath}
\usepackage{amssymb}
\usepackage{mathrsfs}
\usepackage{bbm}
\usepackage{mathtools}
\usepackage{graphicx}
\usepackage{amsthm}	
\usepackage{hyperref}
\usepackage{amsrefs}



\usepackage{babel}

\newtheorem{thm}{Theorem}				

\DeclareMathOperator{\pr}{\mathbf{P}} 		

\begin{document}

\title[The multivariate covering lemma]{The multivariate covering lemma\\ and its converse}

\author{Parham Noorzad}
\address{California Institute of Technology}
\email{parham@caltech.edu}

\author{Michelle Effros}
\address{California Institute of Technology}
\email{effros@caltech.edu}

\author{Michael Langberg}
\address{State University of New York at Buffalo}
\email{mikel@buffalo.edu}

\begin{abstract}
The multivariate covering lemma states that given a collection of $k$
codebooks, each of sufficiently large cardinality
and independently generated according to one of the marginals of a joint distribution, one
can with probability arbitrarily close to one choose one codeword from each codebook such that the 
resulting $k$-tuple of codewords is jointly typical with respect 
to the joint distribution. Prior proofs of the multivariate covering
lemma primarily employ strong typicality. We give a proof of this lemma for 
weakly typical sets. This allows achievability proofs that rely
on the covering lemma to go through for continuous (e.g., Gaussian) channels
 without the need for quantization.
The covering lemma and its converse are widely used in 
information theory, including in rate-distortion theory and in achievability results
for multi-user channels. 
\end{abstract}

\maketitle

\section{Introduction}

The covering lemma and its extensions play a crucial role in achievability 
results in network information theory. 
Covering lemmas are useful for enabling network nodes
to transmit codewords that ``look like'' they are generated
from a dependent distribution, whereas in reality, they are carefully selected
from sufficiently large codebooks that are independently generated. 
This allows nodes to obtain the benefits of both independent and 
dependent codewords: like independent codewords, such 
codewords can be decoded in different locations; 
like dependent codewords they have the potential to achieve rates
higher than those achieved by independent codewords. This benefit, however, comes
at a cost in rate. Thus the strategy is useful when the benefit
transmitting dependent codewords exceeds its cost. 

In the context of the covering lemma, the concept of ``looking like'' dependent codewords
is captured by the notion of being jointly typical with respect to a dependent distribution.
As there are various ways to define the typical set (here we specifically focus
on weakly typical \cite{CoverThomas} and strongly typical sets \cite{ElGamalKim}), one
may ask whether a specific version of the covering lemma holds 
for a given definition of the typical set. The
weakly typical set has two advantages over the strongly typical set. First, it is easily defined
for continuous (e.g., Gaussian) distributions. Second, the weakly typical set has a simple one-shot
counterpart, which allows proofs using the weakly typical set to be written in the one-shot framework
in a simple manner. On the other hand, some results hold for the strongly typical set that do
not hold for the weakly typical set. Thus it is helpful to
review the covering lemma and its extensions and see for which definition of the 
typical set each result is currently known to hold. 

The simplest case of the covering lemma is the situation where given a random vector and an independently
generated codebook, a node looks for a codeword in the codebook that is jointly typical 
(with respect to a dependent distribution) with the given random vector. The result 
obtained in this case, simply referred to as the ``covering lemma'', appears in the
achievability proof of the rate distortion theorem using weakly typical sets
\cite{CoverThomas}. The second case, called the 
``mutual covering lemma,'' treats the case where given two independently 
generated codebooks, a node looks for a jointly typical pair of codewords, where each codeword
is from one of the codebooks. This result is used in Marton's inner bound for the two-user
broadcast channel and is proved for strongly typical sets \cite{Marton, ElGamalMeulen}. 
Recently, by extending the proof of \cite{CoverThomas}, the authors of \cite{Verdu, LiuEtAl} prove
a one-shot version of the mutual covering lemma. This proof can be used to show the validity of
the mutual covering lemma for weakly typical sets in the asymptotic setting.
The proof in \cite{Verdu, LiuEtAl}, however, requires stronger independence assumptions on
the codebooks than the proof using strongly typical sets in \cite{ElGamalMeulen, ElGamalKim}.
Finally, the ``multivariate covering lemma'' is the extension of the mutual
covering lemma to $k$ independently generated codebooks, and can be used to obtain an
inner bound on the broadcast channel with $k$ users \cite{ElGamalKim}. 
As stated in \cite{ElGamalKim}, one can show this result holds for strongly typical 
sets by extending the proof of the mutual covering lemma \cite{ElGamalMeulen}.
 
In this work, using the general strategy of El Gamal and Van der Meulen \cite{ElGamalMeulen}
and some ideas regarding weakly typical sets from Koetter, Effros, and M\'{e}dard \cite{KoetterEtAl2},
we give a proof of the multivariate covering lemma for weakly typical sets. We also 
provide a converse, a special case of which is usually referred to as 
the packing lemma \cite{ElGamalKim}. We remark that while similar to the
argument in \cite{ElGamalMeulen}, we use
Chebyshev's inequality for the direct result (Section \ref{sec:ub}), 
it is also possible to use the Cauchy-Schwarz inequality (see Appendix \ref{app:CSIneq}),
which leads to a more accurate upper bound.

\section{Problem Statement}

For every positive integer $n$, define the set $[n]=\{1,\dots,n\}$.
Let $k$ be a positive integer and 
\begin{equation*}
  p(u_0,u_1,\dots,u_k,u_{k+1})
\end{equation*}
be a probability distribution on the set 
\begin{equation*}
  \prod_{j=0}^{k+1} \mathcal{U}_j.
\end{equation*}
For every nonempty $S\subseteq [k]$ define 
\begin{equation*}
  \mathcal{U}_S=\prod_{j\in S}\mathcal{U}_j.
\end{equation*}
For every $j\in [k]$, let $M_j$ be a nonnegative integer. 
For every nonempty $S\subseteq [k]$, define the set $\mathcal{M}_S$ as
\begin{equation*}
  \mathcal{M}_S=\prod_{j\in S} [M_j].
\end{equation*}
and let $\mathcal{M}=\mathcal{M}_{[k]}$.
For every $\mathbf{m}=(m_1,\dots,m_k)\in \mathcal{M}$,
let the random vector
\begin{equation*}
  (U_0,U_1(m_1),\dots,U_k(m_k),U_{k+1})
\end{equation*} 
have distribution 
\begin{equation*}
  p(u_0)\prod_{j=1}^{k+1} p(u_j|u_0),
\end{equation*} 
where $p(u_0)$ and each $p(u_j|u_0)$ are the conditional 
marginals of $p(u_0,\dots,u_{k+1})$. 
In addition, let $\mathcal{F}$ be an arbitrary subset of 
$\mathcal{U}_0\times\mathcal{U}_{[k+1]}$. We
want to find upper and lower bounds on the probability
\begin{equation*}
 \pr\Big\{\forall \mathbf{m}\in \mathcal{M}:
	\big(U_0,U_1(m_1),\dots,U_k(m_k),U_{k+1}\big)\notin \mathcal{F}\Big\}.
\end{equation*}
We derive the lower bound (Section \ref{sec:lb}) using the union bound, 
which does not depend on the statistical dependencies of the vectors 
\begin{equation*}
  \big(U_0,U_1(m_1),\dots,U_k(m_k),U_{k+1}\big)
\end{equation*}	
for different values of $\mathbf{m}$. For the upper bound (Section \ref{sec:ub}), 
which leads to the multivariate covering lemma, we require a stronger assumption, 
which we next describe.
  
Let $\mathbf{m}=(m_j)_{j\in [k]}$ and $\mathbf{m}'=(m'_j)_{j\in [k]}$ be in $\mathcal{M}$. 
Define the set $S_{\mathbf{m},\mathbf{m}'}$ as
\begin{equation*}
  S_{\mathbf{m},\mathbf{m}'}=\big\{j\in [k]:m_j=m'_j\big\}.
\end{equation*}
When $\mathbf{m}$ and $\mathbf{m}'$ are clear from context, we denote 
$S_{\mathbf{m},\mathbf{m}'}$ with $S$. In the proof of the upper bound
we require 
\begin{align*}
  \MoveEqLeft
  \pr\Big\{\forall j\in [k]:U_j(m_j)=u_j\text{ and }U_j(m'_j)=u'_j\big|
  U_0=u_0,U_{k+1}=u_{k+1}\Big\}\\
  &= \prod_{j=1}^k p(u_j|u_0)\times \prod_{j\in S^c}p(u'_j|u_0),
\end{align*}
for all $u_0$ and
all $(u_j)_j$ and $(u'_j)_j$ such that if $j\in S$, then $u_j=u'_j$ (Assumption I). Note 
that if there exists a $j\in S$ where $u_j\neq u'_j$ then the probability on the 
left hand side equals zero.

In the corresponding asymptotic problem (Section \ref{sec:asymp}), we apply our bounds
to 
\begin{equation*}
  \pr\Big\{\forall \mathbf{m}:
	\big(U_0^n,U_1^n(m_1),\dots,U_k^n(m_k),
	U_{k+1}^n\big)\notin A_\delta^{(n)}\Big\},
\end{equation*} 
where for every $\mathbf{m}$,
\begin{equation*}
  \big(U_0^n,U_1^n(m_1),\dots,U_k^n(m_k),
  U_{k+1}^n\big)
\end{equation*}	
is simply $n$ i.i.d.\ copies of the original random vector
\begin{equation*}
  \big(U_0,U_1(m_1),\dots,U_k(m_k),
  U_{k+1}\big),
\end{equation*}	
(Assumption II) and $A_\delta^{(n)}$ is the weakly typical set for the distribution
$p(u_0,u_1,\dots,u_k,u_{k+1})$. Our main result follows. 

\begin{thm}[Multivariate Covering Lemma] \label{thm:main}
Suppose Assumptions (I) and (II) hold for the joint distribution of
\begin{equation*}
  U_0^n,\big\{U_1^n(m_1),\dots,U_k^n(m_k)\big\}_\mathbf{m},U_{k+1}^n.
\end{equation*}
For the direct part, suppose for all $j\in [k]$, $M_j\geq e^{nR_j}$. If  
for all nonempty $S\subseteq [k]$,
\begin{equation} \label{eq:sumrj}
  \sum_{j\in S} R_j >
  \sum_{j\in S}H(U_j|U_0)-H(U_S|U_0,U_{k+1})+(8k-2|S|+10)\delta,
\end{equation}
then
\begin{equation} \label{eq:limit}
  \lim_{n\rightarrow \infty}
	\pr\Big\{\exists \mathbf{m}:\big(U_0^n,U_1^n(m_1),\dots,U_k^n(m_k),U_{k+1}^n\big)
	\in A_\delta^{(n)}\Big\}=1.
\end{equation}
For the converse, assume for all $j\in [k]$, $M_j\leq e^{nR_j}$. 
If Equation (\ref{eq:limit}) holds, then
\begin{equation*}
  \sum_{j\in S} R_j \geq 
  \sum_{j\in S}H(U_j|U_0)-H(U_S|U_0,U_{k+1})-2(|S|+1)\delta,
\end{equation*}
for all nonempty $S\subseteq [k]$. 
\end{thm}

In the direct part of Theorem \ref{thm:main}, we can weaken the lower bound
on $\sum_{j\in S} R_j$ when $S=[k]$. Specifically, 
we can replace Equation (\ref{eq:sumrj}) with 
\begin{equation*}
  \sum_{j=1}^k R_j > \sum_{j=1}^k H(U_j|U_0)-H(U_{[k]}|U_0,U_{k+1})+2(k+1)\delta.
\end{equation*}
for $S=[k]$. 

\section{The Lower Bound} \label{sec:lb}

For every $S\subseteq [k]$,
define $\mathcal{F}_S$ as the projection of $\mathcal{F}$ on
$\mathcal{U}_0\times\mathcal{U}_S\times\mathcal{U}_{k+1}$.
Then for every $(u_0,u_S,u_{k+1})\in\mathcal{F}_S$, 
let $\mathcal{F}(u_0,u_S,u_{k+1})$ be the set of all $u_{S^c}$ 
such that $(u_0,u_{[k]},u_{k+1})\in\mathcal{F}$.
In addition, for every nonempty $S\subseteq [k]$, let $\alpha_S$ and $\beta_S$ 
be constants such that
\begin{equation*}
  \alpha_S \leq 
  \log \frac{p(u_S|u_0,u_{k+1})}{\prod_{j\in S}p(u_j|u_0)}
\end{equation*}
for all $(u_0,u_S,u_{k+1})\in \mathcal{F}_S$ and 
\begin{equation*}
  \beta_S \leq \log 
	\frac{p(u_S|u_0,u_{S^c},u_{k+1})}{\prod_{j\in S}p(u_j|u_0)}	
\end{equation*}
for all $(u_0,u_S,u_{S^c},u_{k+1})\in \mathcal{F}$. Furthermore, let the
constant $\gamma$ satisfy
\begin{equation*}
  \gamma \geq \log \frac{p(u_{[k]}|u_0,u_{k+1})}{\prod_{j\in [k]}p(u_j|u_0)}
\end{equation*}
for all $(u_0,u_{[k]},u_{k+1})\in\mathcal{F}$. 

For every $\mathbf{m}=(m_1,\dots,m_k)\in \mathcal{M}$, define the 
random variable $Z_\mathbf{m}$ as
\begin{equation*}
  Z_\mathbf{m}=\mathbf{1}\Big\{\big(U_0,U_1(m_1),\dots,U_k(m_k),
  U_{k+1}\big)\in \mathcal{F}\Big\}
\end{equation*}
and set 
\begin{equation*}
  Z=\sum_{\mathbf{m}\in\mathcal{M}}Z_\mathbf{m}.
\end{equation*}
Our aim is to find a lower bound for $\pr\{Z=0\}$. 
Note that for every nonempty $S\subseteq [k]$, 
\begin{align*}
  \pr\big\{\exists \mathbf{m}:Z_\mathbf{m}=1\big\}
	&= \pr\Big\{\exists \mathbf{m}: \big(U_0,U_1(m_1),\dots,U_k(m_k),U_{k+1}\big)\in \mathcal{F}\Big\}\\
	&\leq \pr\Big\{\exists \mathbf{m}: \big(U_0,\big(U_j(m_j)\big)_{j\in S},U_{k+1}\big)\in\mathcal{F}_S\Big\}\\
	&\leq |\mathcal{M}_S|\sum_{\mathcal{F}_S}p(u_0,u_{k+1})\prod_{j\in S}p(u_j|u_0)\\
	&\leq |\mathcal{M}_S|e^{-\alpha_S}\sum_{\mathcal{F}_S}p(u_0,u_S,u_{k+1})\\
	&\leq |\mathcal{M}_S|e^{-\alpha_S}.
\end{align*}
Thus 
\begin{align} \label{eq:lb}
  \pr\{Z=0\} &= 1-\pr\big\{\exists \mathbf{m}:Z_\mathbf{m}=1\big\}\notag\\
	&\geq 1-\min_{|S|\neq\emptyset}|\mathcal{M}_S|e^{-\alpha_S}.
\end{align}

\section{The Upper Bound} \label{sec:ub}

In deriving our upper bound on $\pr\{Z=0\}$, we apply conditioning
and Chebyshev's inequality. Thus, the factor 
\begin{equation*}
  \frac{1}{\big(\pr\{\mathcal{F}(u_0,u_{k+1})\}\big)^2}
\end{equation*}
appears, where 
\begin{align*}
  \pr\{\mathcal{F}(u_0,u_{k+1})\} &=
  \pr\big\{U_{[k]}\in \mathcal{F}(u_0,u_{k+1})|U_0=u_0,U_{k+1}=u_{k+1}\big\}\\
  &=\sum_{u_{[k]}\in\mathcal{F}(u_0,u_{k+1})}p(u_{[k]}|u_0,u_{k+1})
\end{align*} 
and $\mathcal{F}(u_0,u_{k+1})$ (Section \ref{sec:lb}) is simply the set of all
$u_{[k]}$'s that satisfy $(u_0,u_{[k]},u_{k+1})\in\mathcal{F}$. 
Thus to get a reasonably accurate upper bound, we require
$\pr\{\mathcal{F}(u_0,u_{k+1})\}$ to be large. However, as we cannot guarantee this
for all $(u_0,u_{k+1})$, we partition the $(u_0,u_{k+1})$ pairs into ``good'' and ``bad'' sets, 
corresponding to large and small values of 
$\pr\{\mathcal{F}(u_0,u_{k+1})\}$, respectively.
The probability of the good set is large when
$\pr\{(U_0,U_{[k]},U_{k+1})\in\mathcal{F}\}$ 
is sufficiently large. To see this,
fix $\epsilon>0$ and following Appendix III of \cite{KoetterEtAl2}, define the
set $\mathcal{G}\subseteq\mathcal{U}_0\times\mathcal{U}_{k+1}$ as
\begin{equation*}
  \mathcal{G}=\big\{(u_0,u_{k+1}):\pr\{\mathcal{F}(u_0,u_{k+1})\}\geq 1-\epsilon\big\},
\end{equation*}
Note that $\mathcal{G}$ is the set of all good $(u_0,u_{k+1})$ pairs as defined
above. We have
\begin{align*}
  \pr\big\{(U_0,U_{[k]},U_{k+1})\in\mathcal{F}\big\} 
  &=\sum_{u_0,u_{k+1}}\sum_{u_{[k]}\in\mathcal{F}(u_0,u_{k+1})}
  p(u_0,u_{k+1})p(u_{[k]}|u_0,u_{k+1})\\
  &= \sum_{u_0,u_{k+1}}p(u_0,u_{k+1})\pr\{\mathcal{F}(u_0,u_{k+1})\}\\
  &\leq (1-\epsilon)\pr\{(U_0,U_{k+1})\notin\mathcal{G}\}+\pr\{(U_0,U_{k+1})\in\mathcal{G}\}\\
  &= 1-\epsilon \pr\{(U_0,U_{k+1})\notin\mathcal{G}\}.
\end{align*}
Thus
\begin{equation} \label{eq:notinV}
  \pr\{(U_0,U_{k+1})\notin\mathcal{G}\}\leq 
  \frac{1}{\epsilon}\pr\big\{(U_0,U_{[k]},U_{k+1})\notin\mathcal{F}\big\}.
\end{equation}
Our aim is to find an upper bound for $\pr\{Z=0\}$. To do this, we write
\begin{align} \label{eq:initialub}
  \pr\{Z=0\} &= \sum_{u_0,u_{k+1}}p(u_0,u_{k+1})\pr\{Z=0|u_0,u_{k+1}\} \notag\\
  &\leq \frac{1}{\epsilon}\pr\big\{(U_0,U_{[k]},U_{k+1})\notin\mathcal{F}\big\}+
  \sum_{(u_0,u_{k+1})\in \mathcal{G}}p(u_0,u_{k+1})\pr\{Z=0|u_0,u_{k+1}\},
\end{align}
where the inequality follows from Equation (\ref{eq:notinV}).
Therefore, to find an upper bound on $\pr\{Z=0\}$, it suffices
to find an upper bound on $\pr\{Z=0|U_0=u_0,,U_{k+1}=u_{k+1}\}$ for 
all $(u_0,u_{k+1})\in \mathcal{G}$. Fix $(u_0,u_{k+1})\in \mathcal{G}$.
We use Chebyshev's inequality to find an upper bound on 
$\pr\{Z=0|U_0=u_0,U_{k+1}=u_{k+1}\}$. Thus we 
need to calculate $\mathbb{E}[Z|U_0=u_0,U_{k+1}=u_{k+1}]$ 
and $\mathbb{E}[Z^2|U_0=u_0,U_{k+1}=u_{k+1}]$. For a given $\mathbf{m}$,
from the definition of $\gamma$ (Section \ref{sec:lb}) it follows
\begin{align*}
  \mathbb{E}[Z_\mathbf{m}|u_0,u_{k+1}]
  &=\pr\Big\{\big(U_1(m_1),\dots,U_k(m_k)\big)\in 
  \mathcal{F}(u_0,u_{k+1})\big|u_0,u_{k+1}\Big\}\\
  &=\sum_{\mathcal{F}(u_0,u_{k+1})}p(u_1|u_0)\dots p(u_k|u_0)\\
  &\geq \sum_{\mathcal{F}(u_0,u_{k+1})}e^{-\gamma}
  p(u_{[k]}|u_0,u_{k+1})\\
  &= e^{-\gamma}\pr\{\mathcal{F}(u_0,u_{k+1})\}\geq (1-\epsilon)e^{-\gamma}.
\end{align*}
where the last inequality follows from the fact that $(u_0,u_{k+1})\in\mathcal{G}$.
Thus, by linearity of expectation,
\begin{equation} \label{eq:explb}
  \mathbb{E}[Z|U_0=u_0,U_{k+1}=u_{k+1}]\geq |\mathcal{M}|e^{-\gamma}(1-\epsilon).
\end{equation}
Next, we find an upper bound on $\mathbb{E}[Z^2|U_0=u_0,U_{k+1}=u_{k+1}]$. We have
\begin{equation*}
  Z^2 = \sum_{\mathbf{m}}Z_\mathbf{m}^2+
  \sum_{\mathbf{m}\neq\mathbf{m}'} Z_{\mathbf{m}}Z_{\mathbf{m}'}
	=Z+\sum_{\mathbf{m}\neq\mathbf{m}'} Z_{\mathbf{m}}Z_{\mathbf{m}'},
\end{equation*}
since $Z_\mathbf{m}^2=Z_\mathbf{m}$ and $Z=\sum_{\mathbf{m}}Z_\mathbf{m}$.
Thus 
\begin{equation*}
  \mathbb{E}[Z^2|u_0,u_{k+1}]=\mathbb{E}[Z|u_0,u_{k+1}]+\mathbb{E}
	\Big[\sum_{\mathbf{m}\neq\mathbf{m}'} Z_{\mathbf{m}}Z_{\mathbf{m}'}\big|u_0,u_{k+1}\Big]
\end{equation*}
For any pair of distinct $\mathbf{m}$ and $\mathbf{m}'$ with 
nonempty $S=S_{\mathbf{m},\mathbf{m}'}$, we have
\begin{align*}
  \MoveEqLeft
  \mathbb{E}\big[Z_{\mathbf{m}}Z_{\mathbf{m}'}|u_0,u_{k+1}\big]\\
  &=\sum_{\mathcal{F}_S(u_0,u_{k+1})}\prod_{i\in S}p(u_i|u_0)\Big(
  \sum_{u_{S^c}\in \mathcal{F}(u_0,u_S,u_{k+1})}\prod_{j\in S^c}p(u_j|u_0)\Big)^2\\
  &\leq e^{-\alpha_S-2\beta_{S^c}}
  \sum_{\mathcal{F}_S(u_0,u_{k+1})}p(u_S|u_0,u_{k+1})\Big(
  \sum_{u_{S^c}\in \mathcal{F}(u_0,u_S,u_{k+1})}p(u_{S^c}|u_0,u_S,u_{k+1})\Big)^2\\
  &\leq e^{-\alpha_S-2\beta_{S^c}},
\end{align*}
where $\mathcal{F}_S(u_0,u_{k+1})$ is the set of all $u_S$ that satisfy 
$(u_0,u_S,u_{k+1})\in \mathcal{F}_S$.
On the other hand, if $S=S_{\mathbf{m},\mathbf{m}'}$ is empty, then
$Z_\mathbf{m}$ and $Z_\mathbf{m}'$ are independent given 
$(U_0,U_{k+1})=(u_0,u_{k+1})$, and  
\begin{equation*}
  \mathbb{E}\big[Z_{\mathbf{m}}Z_{\mathbf{m}'}|u_0,u_{k+1}\big]
	= \big(\mathbb{E}[Z_\mathbf{m}|u_0,u_{k+1}]\big)^2.
\end{equation*}
Thus (assume $|\mathcal{M}_\emptyset|=1$)
\begin{align} \label{eq:varub}
  \mathbb{E}[Z^2|u_0,u_{k+1}] &= \mathbb{E}[Z|u_0,u_{k+1}]
  +\sum_{S\subset [k]} |\mathcal{M}_S| \prod_{j\in S^c}
  \big(|\mathcal{M}_j|^2-|\mathcal{M}_j|\big)\mathbb{E}[Z_{\mathbf{m}}Z_{\mathbf{m}'}|u_0,u_{k+1}]\notag\\
  &\leq \mathbb{E}[Z|u_0,u_{k+1}]+\big(\mathbb{E}[Z|u_0,u_{k+1}]\big)^2+
  \sum_{\emptyset\subset S\subset [k]} 
  |\mathcal{M}_S||\mathcal{M}_{S^c}|^2 e^{-\alpha_S-2\beta_{S^c}},
\end{align}
where the notation $\emptyset\subset S\subset [k]$ means that $S$ is a nonempty proper
subset of $[k]$. We have
\begin{align*}
 \pr\big\{Z=0|u_0,u_{k+1}\big\}&\leq 
 \pr\Big\{\big|Z-\mathbb{E}[Z|u_0,u_{k+1}]\big|
 \geq \mathbb{E}[Z|u_0,u_{k+1}]\Big|u_0,u_{k+1}\Big\}\\
 &\overset{(a)}{\leq} \frac{\mathrm{Var}(Z|u_0,u_{k+1})}{\big(\mathbb{E}[Z|u_0,u_{k+1}]\big)^2}=
 \frac{\mathbb{E}[Z^2|u_0,u_{k+1}]}{\big(\mathbb{E}[Z|u_0,u_{k+1}]\big)^2}-1\\
 &\overset{(b)}{\leq} \frac{1}{1-\epsilon} |\mathcal{M}|^{-1}e^\gamma
 +\frac{1}{(1-\epsilon)^2}\sum_{\emptyset\subset S\subset [k]}|\mathcal{M}_S|^{-1}
 e^{-\alpha_S-2\beta_{S^c}+2\gamma},
\end{align*}
where (a) follows from Chebyshev's inequality and
(b) follows from Equations (\ref{eq:explb}) and (\ref{eq:varub}). 
Now using Equation (\ref{eq:initialub}), we get
\begin{equation} \label{eq:ub}
  \pr\{Z=0\}\leq \frac{1}{\epsilon}\pr\{\mathcal{F}^c\}+
	\frac{1}{1-\epsilon}|\mathcal{M}|^{-1}e^\gamma
  +\frac{1}{(1-\epsilon)^2}
 \sum_{\emptyset\subset S\subset [k]}|\mathcal{M}_S|^{-1}
 e^{-\alpha_S-2\beta_{S^c}+2\gamma}.
\end{equation}

\section{The Asymptotic Result} \label{sec:asymp}

In this section, using our lower and upper bounds, we prove Theorem \ref{thm:main}.
We first prove the direct part using our upper 
bound from Section \ref{sec:ub}. 
Set $\mathcal{F}=A_\delta^{(n)}$ and for every $j\in [k]$,
choose an integer $M_j\geq e^{nR_j}$. 
Choose a sequence $\{\epsilon_n\}_n$ such
that
\begin{equation*}
  \lim_{n\rightarrow\infty}
  \frac{1}{\epsilon_n}\pr\big\{(A_\delta^{(n)})^c\big\}=0.
\end{equation*}
This is simple to do, since $\pr\big\{(A_\delta^{(n)})^c\big\}$ 
decays exponentially in $n$ (see Appendix \ref{app:largeDev}). 
Fix a nonempty $S\subseteq [k]$. 
Notice that if
$\big(U_0^n,(U_j^n)_{j\in S},U_{k+1}^n\big)\in \mathcal{F}_S$, then
\begin{equation*}
  \Big|\log\frac{p(u_S^n|u_0^n,u_{k+1}^n)}{\prod_{j\in S}p(u_j^n|u_0^n)}
	-n\Big(\sum_{j\in S}H(U_j|U_0)-H(U_S|U_0,U_{k+1})\Big)\Big|
	\leq 2n(|S|+1)\delta.
\end{equation*}
Thus we may choose 
\begin{equation*}
  \alpha_S = n\Big(\sum_{j\in S}H(U_j|U_0)-H(U_S|U_0,U_{k+1})-2(|S|+1)\delta\Big)
\end{equation*}
and 
\begin{equation*}
  \gamma = n\Big(\sum_{j=1}^k H(U_j|U_0)-H(U_{[k]}|U_0,U_{k+1})+2(k+1)\delta\Big).
\end{equation*}
Similarly, for every nonempty $S\subseteq [k]$, we choose 
$\beta_{S}$ as
\begin{equation*}
  \beta_{S} = n\Big(\sum_{j\in S}H(U_j|U_0)-H(U_S|U_0,U_{S^c},U_{k+1})
  -2(|S|+1)\delta)\Big),
\end{equation*}
since for every $\big(U_0^n,(U_j^n)_{j\in S},(U_j^n)_{j\in S^c}\big)\in \mathcal{F}$,
\begin{equation*}
  \Big|\log\frac{p(u_S^n|u_0^n,u_{S^c}^n,u_{k+1}^n)}{\prod_{j\in S}p(u_j^n|u_0^n)}
	-n\Big(\sum_{j\in S}H(U_j|U_0)-H(U_S|U_0,U_{S^c},U_{k+1})\Big)\Big|
	\leq 2n(|S|+1)\delta.
\end{equation*}
From our upper bound, Equation (\ref{eq:ub}), it now follows that if 
for all nonempty $S\subset [k]$,
\begin{align*}
  \sum_{j\in S} R_j &> \frac{1}{n}(2\gamma-\alpha_S-2\beta_{S^c})\\
	&= 2\sum_{j=1}^k H(U_j|U_0)-2H(U_{[k]}|U_0,U_{k+1})
	-\sum_{j\in S}H(U_j|U_0)+H(U_S|U_0,U_{k+1})\\
	&\phantom{=}-2\sum_{j\in S^c}H(U_j|U_0)+2H(U_{S^c}|U_0,U_S,U_{k+1})
  +(8k-2|S|+10)\delta\\
	&=\sum_{j\in S}H(U_j|U_0)-H(U_S|U_0,U_{k+1})+(8k-2|S|+10)\delta,
\end{align*}
and for $S=[k]$,
\begin{equation*}
  \sum_{j=1}^k R_j > \frac{1}{n}\gamma=\sum_{j=1}^k H(U_j|U_0)
  -H(U_{[k]}|U_0,U_{k+1})-2(k+1)\delta,
\end{equation*}
then  
\begin{equation} \label{eq:limit4}
  \lim_{n\rightarrow \infty}
	\pr\Big\{\exists \mathbf{m}:\big(U_0^n,U_1^n(m_1),\dots,U_k^n(m_k),U_{k+1}^n\big)
	\in A_\delta^{(n)}\Big\}=1.
\end{equation}

Next we prove the converse. Suppose for each $j\in [k]$,
$M_j\leq e^{nR_j}$ and Equation (\ref{eq:limit4}) holds. Then
from our lower bound, Equation (\ref{eq:lb}), it follows
\begin{equation*}
  \sum_{j\in S} R_j \geq \frac{1}{n}\alpha_S
	= \sum_{j\in S}H(U_j|U_0)-H(U_S|U_0,U_{k+1})-2(|S|+1)\delta,
\end{equation*}
for all nonempty $S\subseteq [k]$. 

\appendix
\section{Cauchy-Schwarz Inequality} \label{app:CSIneq}

Let $Z$ be any random variable that is nonnegative
with probability one and has positive first and second
moments. Then 
\begin{equation*}
  Z=Z\mathbf{1}\{Z>0\}
\end{equation*}
almost surely. Thus
\begin{align*}
  \mathbb{E}[Z] &= \mathbb{E}\big[Z\mathbf{1}\{Z>0\}\big]\\
  &\leq \sqrt{\mathbb{E}[Z^2]\times \pr\{Z>0\}},
\end{align*}
where the inequality follows from Cauchy-Schwarz. Hence
\begin{equation*}
  \pr\{Z>0\}\geq \frac{\big(\mathbb{E}[Z]\big)^2}{\mathbb{E}[Z^2]}
\end{equation*}
and 
\begin{equation*}
  \pr\{Z=0\}\leq 1-\frac{\big(\mathbb{E}[Z]\big)^2}{\mathbb{E}[Z^2]}.
\end{equation*}
On the other hand, using Chebyshev's inequality we get
\begin{align*}
  \pr\{Z=0\} &= \pr\big\{|Z-\mathbb{E}[Z]|\geq \mathbb{E}[Z]\big\}\\
  &\leq \frac{\mathrm{Var}(Z)}{\big(\mathbb{E}[Z]\big)^2}
  =\frac{\mathbb{E}[Z^2]}{\big(\mathbb{E}[Z]\big)^2}-1.
\end{align*}
Now note that the bound resulting from Cauchy-Schwarz is stronger,
since for any $t>0$,
\begin{equation*}
  1-t\leq \frac{1}{t}-1.
\end{equation*}

\section{Large Deviations} \label{app:largeDev}

The moment generating function of a random variable $X$ is defined as
\begin{equation*}
  M(t)=\mathbb{E}[e^{tX}]
\end{equation*}
for all real $t$ for which the expectation on the right hand side
is finite. If $M$ is defined on a neighborhood of $0$, say 
$(-t_0,t_0)$ for some $t_0>0$, then it has a Taylor series expansion
with a positive radius of convergence \cite[pp. 278-280]{Billingsley}. In 
particular, 
\begin{equation*}
  \frac{d}{dt}M(t)\big|_{t=0}=\mathbb{E}[X].
\end{equation*}

We want to find an upper bound for $\pr\{X\geq a\}$ for
some real number $a$. Choose $t>0$. Using Markov's inequality,
we get
\begin{align*}
  \pr\{X\geq a\} &= \pr\{tX\geq ta\}\\
	&=\pr\{e^{tX}\geq e^{ta}\}\\
	&\leq e^{-ta}\mathbb{E}[e^{tX}]\\
	&= e^{\log M(t)-ta}
\end{align*}
Since $t>0$ was arbitrary, we get
\begin{equation*}
  \pr\{X\geq a\}\leq e^{\inf_{t>0}(\log M(t)-ta)}.
\end{equation*} 
Define the function $f$ as
\begin{equation*}
  f(t)=\log M(t)-ta.
\end{equation*}
Then $f(0)=0$ and $f'(0)=\mathbb{E}[X]-a$. Thus if
$a>\mathbb{E}[X]$, 
\begin{equation} \label{eq:legendre}
  \inf_{t>0}\big(\log M(t)-ta\big)<0.
\end{equation}
If we apply the same inequality to the random variable
\begin{equation*}
  \frac{1}{n}\sum_{i=1}^n X_i,
\end{equation*}
where the $X_i$'s are i.i.d.\ copies of $X$, we get 
\begin{equation} \label{eq:largeDev}
  \pr\Big\{\sum_{i=1}^n X_i\geq na\Big\}\leq e^{n\inf_{t>0}(\log M(t)-ta)}.
\end{equation} 

Now consider a random vector $(U_1,\dots,U_k)$ with distribution 
$p(u_1,\dots,u_k)$. For every nonempty $S\subseteq [k]$, let
$U_S$ denote the random vector $(U_j)_{j\in S}$. 
Let $(U_1^n,\dots,U_k^n)$ be $n$ i.i.d.\ copies of $(U_1,\dots,U_k)$. By applying
inequality (\ref{eq:largeDev}) to the random variables 
$\{\log\frac{1}{p(U_{Si})}\}_{i=1}^n$ and setting $a=H(U_S)+\epsilon$ for 
some $\epsilon>0$, we get
\begin{equation}
  \pr\Bigg\{\sum_{i=1}^n \log\frac{1}{p(U_{Si})}\geq n(H(U_S)+\epsilon)\Bigg\}
	\leq e^{-nI_S(\epsilon)},
\end{equation} 
where $I_S(\epsilon)$ is given by
\begin{equation*}
  I_S(\epsilon)=\inf_{t>0}
	\Big\{t\big(H(U_S)+\epsilon\big)-\log\mathbb{E}\big[p(U_S)^{-t}\big]\Big\}
\end{equation*}
By the union bound we get 
\begin{align*}
  \pr\big\{(U_1^n,\dots,U_k^n)\notin 
	A_\epsilon^{(n)}(U_1,\dots,U_k)\big\}
	&\leq 2\sum_{\emptyset\subsetneq S\subseteq [k]}e^{-nI_S(\epsilon)}\\
	&\leq 2(2^k-1)e^{-n\min_S I_S(\epsilon)}\\
	&\leq e^{-nI(\epsilon)},
\end{align*}
where 
\begin{equation*}
  I(\epsilon)=\min_{S\subseteq [k]}I_S(\epsilon)+o\big(\frac{1}{n}\big).
\end{equation*}
Finally, note that by Equation (\ref{eq:legendre}), each $I_S(\epsilon)$
is positive, thus so is $I(\epsilon)$.

\bibliography{ref}

\end{document}